\documentclass[11pt]{article}
\usepackage{epsfig,cite,graphics}
\def\title{\begin{center}\Large\bf}
\def\author(s){\vspace{0.3cm}\large\rm}
\def\text{\end{center}}

\begin{document}



\noindent {\small{\it Submitted to the Proceedings of the
International Conference dedicated to the 30th anniversary of the
Rozhen Observatory, 27-30 September 2011. Bulgarian Astronomical
Journal 17, 2011}}

\medskip \hrule \medskip

\bigskip


\title
The Synchronous Network of distant Telescopes

\bigskip



\author(s)

B. Zhilyaev$^{ 1}$, O. Svyatogorov$^{ 1}$, I. Verlyuk$^{ 1}$, M.
Andreev$^{ 2}$, A. Sergeev$^{ 2}$, \\ M. Lovkaya$^{ 3}$, S. Antov$^{
4}$, R. Konstantinova-Antova$^{ 4}$, R. Bogdanovski$^{ 5,6}$,
\\ S. Avgoloupis$^{7}$, J. Seiradakis$^{ 7}$, and M. Contadakis$^{ 8}$
\bigskip

\smallskip

\noindent $^1$ {\small {\it Main Astronomical Observatory, NAS
of Ukraine, Kiev, Ukraine}} \\

\noindent {\small {\it e-mail:}} {\small{\bf zhilyaev@mao.kiev.ua}}

\smallskip

\noindent $^2${\small {\it International Centre for Astronomical,
Medical and Ecological Research, Terskol, Kabardino-Balkaria, Russia}} \\

\smallskip

\noindent $^3${\small {\it Crimean Astrophysical Observatory, Nauchny, Crimea, Ukraine }} \\

\smallskip

\noindent $^4${\small {\it Institute of Astronomy and Rozhen NAO, Bulgaria }} \\

\smallskip

\noindent $^5${\small {\it Space Research Institute, Bulgarian Academy of Sciences, Sofia, Bulgaria }} \\

\smallskip

\noindent $^6${\small {\it Institute of Mathematics and Informatics, Sofia, Bulgaria  }} \\

\smallskip

\noindent $^7${\small {\it University of Thessaloniki, Departament
of Physics, Thessaloniki, Greece }} \\

\smallskip

\noindent $^8${\small {\it University of Thessaloniki, Departament of Geodesy and
Surveying, Thessaloniki, Greece }} \\

\smallskip



\text


\section*{Abstract}

The Synchronous Network of distant Telescopes (SNT) represents an
innovative approach in observational astrophysics. Authors present
the unique existing realization of the SNT-conception. It was
founded within the international collaboration between astronomical
observatories of Ukraine, Russia, Bulgaria and Greece. All the
telescopes of the Network are equipped with standardized photometric
systems (based on photomultipliers). The unified timing systems
(based on GPS-receivers) synchronize all the apertures to UTC with
an accuracy of 1 microsecond and better. The essential parts of the
SNT are the original software for operating and data processing.
Described international Network successfully works for more then 10
years. The obtained unique observational data made it possible to
discover new fine-scale features and flare-triggered phenomena in
flaring red dwarfs, as well as the recently found high-frequency
variability in some chromospherically active stars.

\vspace{0.5 cm}


\noindent {{\bf keywords}$\,\,\,\,$\large stars: flare -- stars:
individual: EV Lac -- stars: activity -- methods: observational --
techniques: photometric}

\large

\section{Introduction}

The basic idea of the SNT-conception is to consider a number of
optic telescopes as a single instrument. Such an instrument provides
an elegant solution of wide scope of problems, which remain
irresolvable for single-site observations. Synchronous tracking the
same source with different distant telescopes enables user to obtain
data sets with the same true signal and independent uncorrelated
noise. These obvious advantages enable state-of-the-art experiments,
providing observational information of unprecedented quality.

Authors discuss the following questions: (a) a new observational
technology with the Synchronous Network of distant Telescopes and
new methodology of data processing; (b) description of the optical
High Frequency Oscillations (HFO) in EV Lac flares; (c) the
technique and results of spectroscopic monitoring of transient
events with low-resolution grism spectrometer; etc.


\section{Observations}

\noindent The Synchronous Network of Telescopes, involving
telescopes at four observatories in Ukraine, Russia, Greece and
Bulgaria allows studying small-scale activity of stars with the high
time resolution (Zhilyaev et al. 2003). The instruments used are the
following: the 2 m Ritchey-Chretien and the 60-cm Cassegrain
telescopes at Peak Terskol (North Caucasus, 3100 m a.s.l.) with a
high-speed two-channel UBVR photometer (Zhilyaev et al. 1992); the
30 inch telescope at Stephanion Observatory in Greece, equipped with
a single-channel photometer with digitized readings in the U band
(Mavridis et al. 1982); the 1.25 m reflector AZT-11 at the Crimean
Observatory with a UBVRI photometer-polarimeter (Kalmin \&
Shakhovskoy 1995); the 2 m Ritchey-Chretien telescope at the Rozhen
Observatory and the 60-cm Cassegrain telescope at the Belogradchik
Observatory with a single-channel UBV photon-counting photometer
(Antov \& Konstantinova-Antova 1995).  All the telescopes of SNT are
equipped with standardized photometric systems based on
photomultipliers. The unified timing systems based on GPS technology
synchronize all the apertures to UTC with an accuracy of 1
microsecond and better. For operating and processing of the SNT data
the original software was developed. The typical integration time is
0.1 s. To increase the signal-to-noise ratio we perform digital
filtering of the available photometric data. We use the Kaiser
convolution to obtain a low-frequency outburst light curve from the
sets of flaring stars data and a moving-average to suppress
high-frequency noises (see details in Zhilyaev et al. 2000).

The SNT team carries out usually two observing campaigns yearly, the
winter and the autumn. Figure 1 represents the light curves of the
flare detected on Sept. 12, 2004. All sites in the Ukraine, Russia,
Greece, and Bulgaria had registered this flare event. Figure 1
demonstrates the remarkable consistency of the U-band light curves
simultaneously obtained by three instruments of the Crimean,
Stephanion, and Belogradchik observatories. The smoothed data show
clear presence of high-frequency oscillations (HFO) in the
descending part of the flare. The application of new observational
technology and new methodology of data processing - a high-speed,
multi-site monitoring and digital filtering - provided new results:
we confirmed HFO in stellar flares and discovered fast color
variations of the flare's radiation (Zhilyaev et al. 2007). The
color-indices and time tracks in the UBVR color-color diagrams are
based on the data obtained with the Crimean 1.25 m reflector AZT-11.
An example of such color time tracks can be seen in Figure 2.

Figure 2 shows the color tracks of the flare from October 10, 1998.
According to the theoretical color-color diagrams (Chalenko 1999),
the flare's onset (and 95\% error ellipse) falls in the region
corresponding to the radiation of hydrogen plasma optically thin in
the Balmer continuum. The flare in its oneset shows extremely blue
colors U-B = -1.1, B-V = -1.3. Using the blackbody model for the
photosphere of EV Lac in a quiet state and the observed flare
amplitudes in the U-band, one may estimate the size of flare
(Alekseev \& Gershberg 1997). For EV Lac in quiescence one may adopt
$T_{eff} = 3300 K$ (Pettersen 1980). Changes in the color-color
diagram prove that the flare in its maximum is similar to a
blackbody source with the most probable temperature of 16 000 K.

\begin{figure}[!htb]
  \begin{center}
    \centering{\epsfig{file=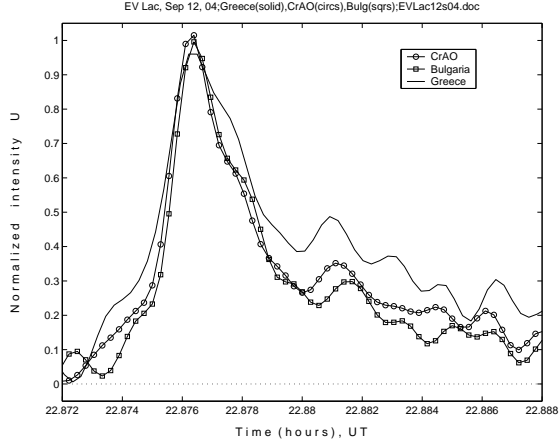, width=0.5\textwidth}}
    \caption[]{Multi-site photometry of a flare on EV Lac from September 12, 2004,
22:53 UT (max), as simultaneously seen by the three instruments:
Ukraine (circles), Greece (solid), and Bulgaria (squares).}
    \label{countryshape}
  \end{center}
\end{figure}

\begin{figure}[!htb]
  \begin{center}
    \centering{\epsfig{file=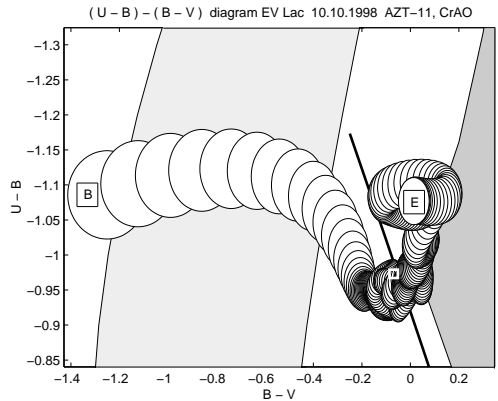, width=0.5\textwidth}}
    \caption[]{Color tracks of a flare on EV Lac from Oct. 10, 1998, as observed from
the Crimean 1.25 m reflector AZT-11. The 95\% error ellipses are
shown. The onset and end of the flare are marked by letters B and E,
respectively. The light density area denotes color characteristics
of plasma optically thin in the Balmer continuum, with $T_{e} \simeq
$ 15000 K and $N_{e}$ from $10^{14}$ to $10^{10}\,cm^{-3}\,$, the
gray density area covers the region corresponding to plasma
optically thick in the Balmer continuum, with $T_{e}$  from 15000 to
8000 K. Blackbody radiation follows the heavy line (Chalenko 1999).}
    \label{countryshape}
  \end{center}
\end{figure}

\section{The grism spectroscopy of transient events}

The slitless UBVR grism spectrometer is established on the Zeiss-600
telescope at Peak Terskol in North Caucasus. It allows to carry out
parallel recording of the spectrum with subsecond temporal
resolution while working the SNT. A blazed transmission grating is
included in the converging beam in the telescope filter wheel. The
wavelength scale after calibration is accurate to about $30{\AA}$.
The grating spectrum has a resolution of $R \approx 100$ at
$4800{\AA} $. The grism spectrometer can provide moderate
signal-to-noise ratio for stars up to 16 magnitude. It allows
measuring the equivalent widths of the non blended lines down to
$0.7 {\AA}$. The grism tools can be used to study temporal
variations in the spectra of variable stars. We use customized
software that relies on the theory of count statistics to identify
intrinsic activity in spectra. This allows us to detect the relative
power of fluctuations down to $(10^{-5} - 10^{-6})$ in the optics.
The obtained unique observational data made it possible to discover
new fine-scale features and flare-triggered phenomena in flaring red
dwarfs, as well as low-amplitude rapid variability of spectra in
chromospherically active stars. Detailed colorimetric analysis based
on the fast grism spectroscopic data allowed estimating important
characteristics of flares on some red dwarf flaring stars: the
temperature at brightness maximum and size. The high temporal
resolution monitoring of some chromospherically active stars
revealed variations in the Balmer and CaII H, K lines at time span
of seconds to minutes with amplitudes of a few percents. This allows
us first to assert the existence of intense microflaring activity in
these stars.

\begin{figure}[!htb]
  \begin{center}
    \centering{\epsfig{file=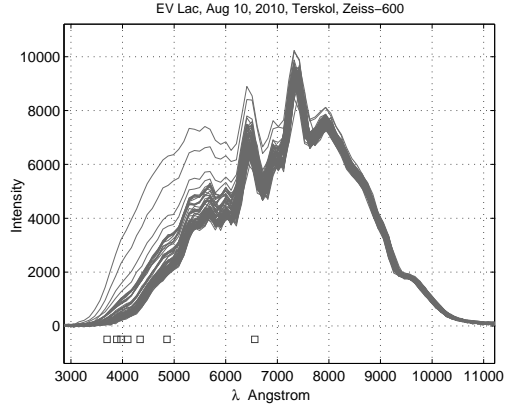, width=0.5\textwidth}}
    \caption[]{Grism spectra of strong flare on EV Lac on August 10,
    2010, obtained  with the grism spectrometer on the 60 cm Zeiss telescope.}
    \label{countryshape}
  \end{center}
\end{figure}

\begin{figure}[!htb]
  \begin{center}
    \centering{\epsfig{file=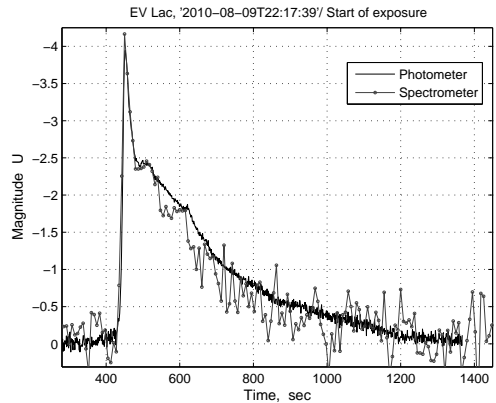, width=0.5\textwidth}}
    \caption[]{The U-band light curves of a flare event on EV Lac, August 10, 2010,
    based on the synchronous observations with the Terskol 2-m telescope
    (two channel photometer), and the 60 cm Zeiss telescopes (grism spectrometer).}
    \label{countryshape}
  \end{center}
\end{figure}

\begin{figure}[!htb]
  \begin{center}
    \centering{\epsfig{file=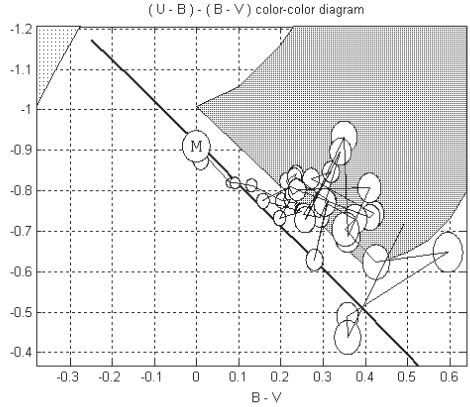, width=0.5\textwidth}}
    \caption[]{Color tracks of a flare in the diagram (U - B) - (B - V) from the grism data.
    Ellipse with the letter M marks the position of the flare maximum. At maximum
    light a flare emits as a blackbody with a temperature of $13400 \pm 500$ K.
    The size of flare is 3.9\% radius of the star, or about 0.15\% of the visible disk of the star.}
    \label{countryshape}
  \end{center}
\end{figure}


\section{Some results}

\subsection{The flare star EV Lac}

100 spectra of strong flare on EV Lac with a time resolution of 7.5
sec are shown in Figure 3. One can easily see a dramatic change in
the short wavelength during a flare. By mathematical convolution of
spectra with the transmission curves of filters one can obtain
estimates of the UBVR magnitudes. The U band light curve of strong
flare of EV Lac on Aug 10, 2010 is shown in Figure 4. The
synchronous registration of the flare with the 2 m telescope
displays in Figure 4 too. Detailed colorimetric analysis allowed
estimating the temperature at maximum brightness and its size. The
two-color (U-B) - (B-V) diagram in Figure 5 confirms that the flare
at maximum light radiates as a black body with a temperature of
about 13 400 K. The linear size of the flare at the maximum
luminosity is about 3.9\% of the stellar radius.

\subsection{The chromospherically active star SAO 52355}

We have carried out low-resolution spectroscopy of SAO 52355 with
the Zeiss-600 telescopes at Peak Terskol. SAO 52355 is a field star
of the spectral type K0 III. We came out on this star in search of
reference stars for the variable EV Lac. The X-ray observations
taken by Ginga and ROSAT indicate in SAO 52355 coronae plasma
temperatures as high as $10^{7}$ K. This star is supposed to have
high-powered chromosphere.

On  May 30,  2010, we obtained 200 low-resolution grating
spectrograms of SAO 52355. For SAO 52355 we used  GSC 3226 640 as
the reference star, Johnson V magnitude: 10.69. Data inferred from
the Tycho catalog: BT magnitude $11.018 \pm 0.032$, VT magnitude
$10.726 \pm 0.036$. Johnson B-V color, computed from BT and VT:
0.270.

In Figure 6 the relative power of variations in grating spectra of
SAO 52355 and its reference star GSC 3226 640 are shown. Note the
"smooth" spectral energy distribution in the power spectra of the
reference star and the "emission" features at wavelengths of the
Balmer lines and the CaII H, K lines, as well as at the wavelength
of the Balmer jump $\lambda \,\, 3700-3800\,{\AA} $ in SAO 52355.
From the power spectrum data one can find that variations in the
intensities of the CaII H, K and $H_{\gamma}$ lines are 3.2\% and
1.5\%, respectively.


\begin{figure}[!htb]
  \begin{center}
    \centering{\epsfig{file=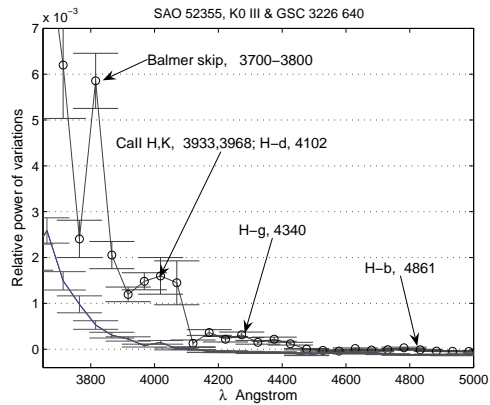, width=0.5\textwidth}}
    \caption[]{The relative power of variations in grating spectra of SAO
52355 (circles) and its reference star  GSC 3226 640 (solid curve).
1 sigma error bars are shown.}
    \label{countryshape}
  \end{center}
\end{figure}

It is found that chromospherically active stars show strong Hydrogen
and Ca II emission lines as well as high X-ray luminosity.  As
mentioned by Gondoin (2003), the high X-ray flux from
chromospherically active stars detected by space observatories could
be explained by assuming that the heating of its corona results from
a large number of small flares.  It is well-known that the stellar
coronae are heated by the two most favored agents involving magnetic
fields, namely MHD waves and transients, such as flares, and
microflares (see Narain \& Pandey 2006, and references therein). It
is thought that flares are associated with conversion of magnetic
energy of plasma to thermal energy via magnetic reconnection. The
reconnect heating by foot-point motions was proposed first by Parker
(1991).

A forest of closed magnetic loops has foot points ankered in
photospheric regions. The mechanical energy flux is generated by
foot point motions. These motions increase the energy stored in the
entwined magnetic field. This system can return to a minimum energy
configuration only after a reconnection (or a cascade of
reconnections). It is thought that these small and frequent
reconnection events give rise to the microflare heating (Narain \&
Ulmschneider 1996).

Thus, our observations allows us to assert the existence of intense
microflaring activity in the chromospherically active star SAO
52355.

\section*{Conclusion}

The application of new observational technology with the Synchronous
Network of distant Telescopes and new methodology of data
processing, provided new results: we confirmed High-Frequency
optical brightness Oscillations (HFO) in stellar flares and
discovered fast color variations of the flare radiation.

Our first description of the optical HFO in the EV Lac flares
(Zhilyaev et al. 2000) was interpreted by Kouprianova et al. (2004)
and Stepanov et al. (2005). They concluded that a source of such an
HFO is located at a magnetic loop footpoint, the pulsations are
determined by modulation of the flux of energetic particles
descending along a loop with fast magnetoacoustic oscillations.

Spectroscopic monitoring of the chromospherically active star SAO
52355 showed variations of intensity in the Balmer lines and in the
CaII H, K lines at time intervals ranging from seconds to minutes.
From the power spectrum data one can find that variations in the
intensities of the CaII H, K and $H_{\gamma}$ lines are 3.2\% and
1.5\%, respectively. Satellite surveys at X-ray indicate high
temperature coronae plasma in many chromospherically active stars.
High-frequency changes, which were found in SAO 52355, suggests the
existence of intense microflaring activity in this star.

\end{document}